\documentclass[12pt,a4paper]{article}

\usepackage[T1]{fontenc}
\usepackage{lmodern}
\usepackage[margin=2.5cm]{geometry}
\usepackage{setspace}
\usepackage{graphicx}
\usepackage{amsmath,amssymb,amsfonts}
\usepackage{array}
\usepackage{xcolor}
\usepackage[authoryear,round]{natbib}
\usepackage{url}
\usepackage[title]{appendix}
\usepackage[hidelinks]{hyperref}

\graphicspath{{figures/}}

\providecommand{\doi}[1]{\href{https://doi.org/#1}{doi:#1}}

\DeclareMathOperator*{\argmin}{arg\,min}
\DeclareMathOperator{\MSE}{MSE}
\newcommand{\E}{\mathrm{E}}
\newcommand{\Var}{\mathrm{Var}}

\newcommand{\noisy}{\mathrm{noisy}}
\newcommand{\zne}{\mathrm{ZNE}}
\newcolumntype{L}[1]{>{\raggedright\arraybackslash}p{#1}}

\title{Decision-aware validation of zero-noise extrapolation in QAOA optimization landscapes}
\author{Vicenzo Scavino Alfaro\\[4pt]
\small Independent Researcher, Lima, Peru\\
\small ORCID: \href{https://orcid.org/0009-0000-2472-9785}{0009-0000-2472-9785}\\
\small E-mail: \href{mailto:vicenzoscavino@hotmail.com}{vicenzoscavino@hotmail.com}}
\date{}

\begin{document}

\maketitle

\begin{abstract}
Zero-noise extrapolation (ZNE) is usually validated by the accuracy of mitigated observables, yet variational quantum optimization uses those observables to select parameters. We ask whether, and by how much, an accuracy gain from ZNE coincides with a worse decision. In depth-one simulations of the quantum approximate optimization algorithm (QAOA) for MaxCut on four to six qubits, each sampled $120\times120$ landscape yields both its mean-squared error (MSE) and the ideal regret of its selected grid minimum, comparing Richardson orders one to three with unmitigated estimation at the same total shot budget per grid point. In an exploratory screen of 675 conditions, whose per-point random seeds collide, a familywise (Holm) analysis retains only four of 540 replication conditions as showing both directional MSE improvement and regret increase. A frozen confirmation with 100 fresh paired repetitions per condition reproduces all four: median MSE improvements of 7.65--73.09\% accompany median normalized regret increases of only 0.0031--0.0073. We distinguish a bias route, in which residual bias reorders the mean landscape, from a variance route, in which Richardson weights amplify finite-shot noise. Exact first and second moments on a bounded control grid find no bias-route reversal: mitigated mean landscapes preserve the ideal minima, and ZNE can more than halve expected MSE, yet finite-shot selections incur higher mean regret. Decision losses that survive strict testing are few and small in this benchmark, but reproducible and not revealed by aggregate MSE. We present a paired benchmark and recommend reporting decision regret alongside estimation error when mitigation is assessed for optimization. Results concern simulated exhaustive grids, not hardware or optimizer loops.
\end{abstract}


\section{Introduction}
Variational quantum optimization, as in the quantum approximate optimization algorithm (QAOA), uses estimated objectives to choose circuit parameters \citep{Farhi2014,Zhou2020QAOA,Cerezo2021VQAReview}. Error mitigation can improve those estimates without full fault tolerance, at a sampling cost that matters when measurements are limited \citep{Temme2017,LiBenjamin2017,Cai2023QEM,Takagi2022LimitsQEM}. Zero-noise extrapolation (ZNE) is usually validated by the accuracy of the mitigated observable. For optimization, however, the quantity of interest is the quality of the parameter that the mitigated estimate selects. Richardson extrapolation reduces bias by combining folded circuits with weights whose magnitudes amplify sampling variance \citep{Krebsbach2022RichardsonOptimization}, and the two error sources need not affect accuracy and decisions in the same way: landscape mean-squared error (MSE) averages over every parameter, whereas the selected parameter is decided by its contrasts with the ideal minimum, which are smallest, and therefore most easily reversed, at near-optimal parameters. Whether an accuracy gain becomes a decision gain is therefore an empirical question for each mitigation setting.

The general distinction is well established. Predict-then-optimize methods separate prediction error from the loss of the decision induced by a prediction \citep{ElmachtoubGrigas2022SPO}. In variational quantum algorithms, convergence analyses with quasiprobabilistic mitigation balance reduced bias against added variance \citep{JoseSimeone2022Convergence}, and mitigation can fail to restore the resolvability of cost differences at acceptable resource cost \citep{Wang2024Trainability}; the QAOA optimizer-loop experiments in that study find no improvement from ZNE over noisy optimization in the evaluated settings. Landscape diagnostics show that Richardson extrapolation can introduce roughness that complicates gradient-based optimization \citep{Liu2023OSCAR}. Shot-budget benchmarks compare mitigated QAOA MaxCut observables at parameters optimized beforehand without noise \citep{Bultrini2023UNITED}, and hardware QAOA studies combine ZNE with readout correction and layout selection while evaluating optimization trajectories and final sampled-solution success \citep{Cheng2024DARBO}. Conference reports examine ZNE in QAOA workflows \citep{Venere2024ZNEQAOAWorkflow,Lusso2025NoiseMitigationQAOA}, and mitigation within variational training has been demonstrated for function fitting \citep{Robbiati2026RealtimeQEMVQA}. Studies of local noise, parameter setting and energy landscapes characterize the underlying QAOA problem \citep{Marshall2020LocalNoiseQAOA,Sureshbabu2024ParameterSettingWeightedQAOA,BoyWales2024EnergyLandscapesQAOA}.

The closest precedents compare estimation quality with downstream outcomes. Sack and Egger report QAOA runs in which optimizing a machine-learning-mitigated cost yields worse parameters under noiseless sampling than optimizing the unmitigated cost \citep{SackEgger2024MLMitigation}; their reported MSE concerns correlator prediction, not whole landscapes. A response-model-learning preprint reports landscape root-mean-square error and selected-point regret separately, using different corrections and calibration budgets from a raw-versus-ZNE comparison \citep{Khaleghian2026QAOAReliability}. A preprint by the present author develops residual-gap theory for finite-shot mitigation, including ZNE, and evaluates static failure probabilities and dynamic regret \citep{Scavino2026DecisionKernels}. What remains unquantified for conventional Richardson ZNE is whether, and by how much, a whole-landscape accuracy gain coincides with a worse selected parameter when both are measured on the same sampled landscape, at an equal shot budget, and without the trajectory and stopping-rule choices of an optimizer.

We measure this paired relationship in depth-one QAOA MaxCut on four to six qubits. Exhaustive $120\times120$ grids replace the optimizer, so each sampled landscape yields both its MSE against the exact ideal reference and the ideal regret of its selected minimum. Mitigated and unmitigated estimates use the same total number of shots at each grid point. Fifteen graphs and 675 conditions, spanning five depolarizing levels and three budgets, form an exploratory screen. Four selected conditions are then confirmed prospectively on fresh data, and an exact-moment control separates squared bias from sampling variance. The contributions are:
\begin{itemize}
\item \emph{A paired accuracy--decision benchmark.} Per-repetition landscape MSE and ideal regret measured on the same sampled landscape, at equal per-point budgets and against exact ideal grids.
\item \emph{Reproducible but small discordance.} A familywise (Holm) analysis of the exploratory screen retains four conditions that combine directional MSE improvement with directional regret increase, and a frozen confirmation on fresh repetitions reproduces all four, with regret shifts below the pre-specified practical-effect threshold.
\item \emph{Bias and variance routes to a decision loss.} Decomposing the observed contrast separates a bias route, in which residual bias reorders the mean landscape, from a variance route, in which Richardson weights amplify finite-shot noise. An exact-moment control finds the variance route at work, and an oracle reallocation of shots shows that the implemented allocation is already close to variance-optimal in the confirmed conditions.
\end{itemize}

The scope is bounded. The screen is exploratory: its sampler reuses some random streams, and its directional analysis was chosen after the screening results were known. The confirmation addresses only the four selected conditions, so it establishes reproducibility rather than frequency across graphs. Hardware, adaptive optimizers and deeper circuits lie outside this simulation study.

We first define the design and paired endpoints, then show why a relative MSE gain cannot order decisions and identify the bias and variance routes. The statistical analysis precedes the confirmation, screening, control and allocation results, followed by a discussion of limitations.

\section{Problem and experimental design}
\subsection{Finite-grid QAOA and paired endpoints}
For a positive-weight graph $G=(V,E,w)$, let
\begin{equation}
 H_C=\sum_{(i,j)\in E}w_{ij}(I-Z_iZ_j)/2,\qquad H_M=\sum_i X_i.
\end{equation}
Depth-one QAOA prepares $|\psi(\gamma,\beta)\rangle=e^{-i\beta H_M}e^{-i\gamma H_C}|+\rangle^{\otimes |V|}$. We minimize $L_0(\gamma,\beta)=-\langle\psi|H_C|\psi\rangle$ over a $120\times120$ grid $\Theta$ on $[0,2\pi)\times[0,\pi)$. Zero is included and each upper endpoint is excluded. This is a prescribed finite domain, and the results concern it alone; arbitrary weighted graphs need not have a $2\pi$ period in $\gamma$. In $\beta$ the depth-one landscape has period $\pi/2$: $e^{-i(\pi/2)H_M}=(-i)^{|V|}X^{\otimes|V|}$, and $X^{\otimes|V|}$ commutes with both $H_M$ and $H_C$, so it leaves $\langle H_C\rangle$ unchanged. The $\beta$ range therefore contains two copies of the landscape, aligned with the grid because $\pi/2$ spans 60 grid steps, and the ideal minimizers on $\Theta$ occur in symmetry-related pairs whose computed values agree up to floating-point roundoff. Such symmetries produce the equivalent ideal minima that the analysis below treats explicitly.

Write $L_*=\min_\Theta L_0$, $D_0=\max_\Theta L_0-L_*>0$, and $\Theta_*=\argmin_\Theta L_0$. A row-major ordering resolves ties in any sampled landscape. For $\widehat\theta=\argmin_\Theta\widehat L$, define
\begin{align}
 \MSE(\widehat L,L_0)&=|\Theta|^{-1}\sum_{\theta\in\Theta}(\widehat L(\theta)-L_0(\theta))^2,\\
 R(\widehat L;L_0)&=\{L_0(\widehat\theta)-L_*\}/D_0.
\end{align}
This is normalized simple regret for a terminal recommendation from a finite set, as in fixed-budget best-arm identification \citep{AudibertBubeckMunos2010BestArm}. Here observations may remain biased relative to the ideal objective and multiple ideal optima are allowed, so standard best-arm guarantees are not imported. For each graph/noise/budget/order condition and sampling seed, the paired quantities are
\begin{align}
 r_{\MSE}&=\frac{\MSE(\widehat L_{\noisy},L_0)-\MSE(\widehat L_{\zne},L_0)}{\MSE(\widehat L_{\noisy},L_0)},\label{eq:rmse}\\
 d_R&=R(\widehat L_{\zne};L_0)-R(\widehat L_{\noisy};L_0).\label{eq:dr}
\end{align}
Positive values of both describe improved aggregate accuracy and a worse selected parameter for that repetition. The denominator in Eq.~\eqref{eq:rmse} is positive in the retained data. Regret uses the exact ideal reference available in simulation.

\subsection{Noise, extrapolation and resource accounting}
Local depolarizing channels follow circuit gates. The channel on a $d$-dimensional subsystem is $\mathcal D_\kappa(\rho)=(1-\kappa)\rho+\kappa I/d$, with $\kappa=p$ after one-qubit gates and $\kappa=2p$ after two-qubit gates. Thus $p$ is the depolarizing-channel parameter; the corresponding probabilities of a nonidentity Pauli error are $3p/4$ after one-qubit gates and $15p/8$ after two-qubit gates. We use $p\in\{0.001,0.003,0.005,0.01,0.02\}$.

Deterministic global circuit folding \citep{GiurgicaTiron2020DigitalZNE} produces scale sets $(1,3)$, $(1,3,5)$ and $(1,3,5,7)$ for Richardson orders $m=1,2,3$. Folding preserves the ideal circuit, but local noisy folded circuits are not assumed to equal a single channel with parameter $a_jp$. The estimator is
\begin{equation}
 \widehat L_Z(\theta)=\sum_j\lambda_j\widehat L^{[a_j]}(\theta),\qquad
 \sum_j\lambda_j=1,\quad\sum_j\lambda_j a_j^r=0\ (1\le r\le m).
\end{equation}
The total budget $B\in\{512,1024,4096\}$ is allocated across scales in proportion to $|\lambda_j|$, using largest remainders to obtain positive integers $B_j$ with $\sum_j B_j=B$. The unmitigated estimate uses all $B$ shots. This is a budget \emph{per grid point}: one full landscape costs $14\,400B$ shots ($14\,745\,600$ at $B=1024$). Each screening condition has 30 sampling repetitions; the same noisy baseline is shared between extrapolation orders for a given graph, noise, budget and seed.

\subsection{Discovery and generated replication}
Discovery comprises three graphs and $3\times5\times3\times3=135$ conditions. Replication comprises twelve additional graphs, H1--H12, and $12\times5\times3\times3=540$ conditions. Graphs have four to six vertices. Two replication graphs were selected in each of six strata: sparse, regular or near-regular, Erdos--Renyi, heterogeneous weighted, dense positive-weight, and hardware-friendly/control. Graph generation, selection and grid-adequacy checks were fixed before any noisy simulation, which protects instance selection from the outcomes.

Probabilities were evaluated with a custom NumPy density-matrix kernel validated against Qiskit Aer within the Qiskit software framework \citep{JavadiAbhari2024}; a bounded cross-check of 540 graph/point/noise/scale combinations against Aer gave a maximum objective discrepancy $9.45\times10^{-14}$. Probabilities are cached across budgets and repetitions, and finite-shot samples use deterministic pseudorandom seeds.

\section{Why relative MSE improvement does not order decisions}\label{sec:mse_decisions}
\subsection{A finite-grid example and an absolute-error guarantee}\label{sec:nonimplication}
On two points, take $L_0=(0,1)$, $\widehat L_N=(2,3)$ and $\widehat L_Z=(0.6,0.4)$. Their MSEs are $4$ and $0.36$, yet their normalized regrets are $0$ and $1$; the landscapes are arbitrary and illustrate only the logical nonimplication. The same construction works on any finite grid $\Theta$ of size $N$ with nonconstant $L_0$, including grids with several ideal minimizers. For a point with ideal gap $\Delta>0$, adding a constant $C>2\Delta/\sqrt N$ to $L_0$ preserves the ideal optimal set, whereas lowering that point by $2\Delta$ makes it the unique observed minimizer, with regret $\Delta/D_0$ and the smaller MSE $4\Delta^2/N<C^2$.

Absolute MSE nevertheless controls regret on a fixed finite grid. With $e=\widehat L-L_0$,
\begin{equation}
 R(\widehat L;L_0)\le\min\left\{1,\frac{\sqrt{2N\MSE(\widehat L,L_0)}}{D_0}\right\}.\label{eq:mse_bound}
\end{equation}
Indeed, for $u\in\Theta_*$, selection implies $L_0(\widehat\theta)-L_*\le e(u)-e(\widehat\theta)$. Cauchy--Schwarz bounds this contrast by $\sqrt{2\sum_\theta e(\theta)^2}$. Likewise $\|e\|_\infty\le\delta$ implies unnormalized regret at most $2\delta$. If the smallest positive ideal gap is $g_*$, then $2\delta<g_*$ ensures that every selected minimizer belongs to $\Theta_*$. Thus MSE tending to zero does imply vanishing regret on a fixed grid. A relative improvement between two finite errors does not order their actual regrets.

\subsection{Integer-budget variance and two routes to a decision change}\label{sec:routes}
If the single-shot objective variance at folded scale $j$ is $v_j(\theta)$, independent samples across scales give
\begin{equation}
 \Var[\widehat L_Z(\theta)]=\sum_j\frac{\lambda_j^2v_j(\theta)}{B_j}
 \le v_{\max}(\theta)S_B,\qquad S_B=\sum_j\frac{\lambda_j^2}{B_j}.\label{eq:integer_variance}
\end{equation}
For an exactly proportional allocation, including the continuous ideal, $S_B=A_m/B$, where $A_m=(\sum_j|\lambda_j|)^2$. The weights and amplification factors are
\begin{equation}
\begin{array}{c|c|c}
m&\lambda&A_m\\\hline
1&(3/2,-1/2)&4\\
2&(15/8,-5/4,3/8)&49/4\\
3&(35/16,-35/16,21/16,-5/16)&36
\end{array}
\end{equation}
Integer rounding makes $S_B\ge A_m/B$, with relative excess at most $1.39\times10^{-5}$ for the nine implemented allocations (Appendix~\ref{app:mathematics}). Proportional allocation minimizes this variance upper-bound coefficient; minimizing the actual variance for unequal $v_j$ instead gives $B_j\propto|\lambda_j|\sqrt{v_j}$ in the continuous positive-variance case \citep{Krebsbach2022RichardsonOptimization}.

Write $\widehat L_Z=L_0+b_Z+\eta_Z$, where $b_Z$ is the exact mean residual and $\eta_Z$ is centered sampling noise. For a fixed ideal minimizer $u$ and competitor $v$, the observed contrast is
\begin{equation}
 \widehat L_Z(v)-\widehat L_Z(u)=L_0(v)-L_*+b_Z(v)-b_Z(u)+\eta_Z(v)-\eta_Z(u).\label{eq:contrast}
\end{equation}
Beating one ideal minimizer is a pairwise ranking event, not a guarantee that $v$ is the selected global minimum. With equivalent optima, a sufficient condition for positive regret is that some nonoptimal point beats \emph{all} ideal minimizers. By Eq.~\eqref{eq:contrast}, $v$ does so whenever, for every $u\in\Theta_*$, $|\eta_Z(v)-\eta_Z(u)|\le\delta_u$ and $b_Z(v)-b_Z(u)+\delta_u<-[L_0(v)-L_*]$; if the unmitigated selection lies in $\Theta_*$, this also gives $d_R>0$.

Equation~\eqref{eq:contrast} separates two routes by which ZNE can change a decision relative to the ideal ordering. In the \emph{bias route}, the residual contrast $b_Z(v)-b_Z(u)$ reverses the order of $v$ and $u$ even without sampling noise. The exact Richardson example in Appendix~\ref{app:mathematics} shows that this route is compatible with lower MSE: on two points with bounded expectation values, Richardson extrapolation lowers the deterministic MSE from 12.5 to 4.5 while reversing a decision that the unmitigated expectation preserves. The example is abstract, and the studied circuits need not realize such coefficients. In the \emph{variance route}, the mean landscape keeps the ideal ordering, but the centered contrast $\eta_Z(v)-\eta_Z(u)$ reverses it in some finite-shot repetitions. By Eq.~\eqref{eq:integer_variance}, the scale of that contrast is governed by $S_B\ge A_m/B$, at least $A_m$ times the coefficient $1/B$ of unmitigated estimation at the same budget. The routes are not exclusive, because residual bias also changes the margins that sampling noise must overcome.

Appendix~\ref{app:mathematics} applies Hoeffding's inequality \citep{Hoeffding1963} to bound the probability of excess regret in terms of $S_B$ and an ideal anchor, without assuming a unique optimum; Section~\ref{sec:discussion} interprets it. It is a structural bound under independent shots rather than a calibrated predictor of $d_R$. Section~\ref{sec:control} examines which route is present in a bounded control.

\section{Statistical analysis}\label{sec:statistics}
\subsection{Screening analysis}\label{sec:screening}
The exploratory screen is analyzed with the directional tests below and Holm correction. This analysis replaces the pre-specified bootstrap classification rule, whose results and calibration and resolution limitations are reported as a secondary descriptive analysis in Appendix~\ref{app:statistics}. Nine discovery and 33 replication conditions excluded by the detection-floor rule, which flags noisy MSE below $10^{-4}D_0^2$ in some seed, remain in their families with $p_c=1$.

\subsection{Directional inference}\label{sec:directional}
For a fixed condition $c$, the reference inferential model assumes independent, identically distributed sampling repetitions with circuit probabilities held fixed. We test
\begin{equation}
 \pi_{r,c}=\Pr(r_{\MSE}>0)>1/2,\qquad
 \pi_{d,c}=\Pr(d_R>\epsilon_{\rm num})>1/2,\quad\epsilon_{\rm num}=10^{-12}.\label{eq:estimand}
\end{equation}
The small regret tolerance treats roundoff-level splitting of equivalent ideal minima as nonpositive; it is not a practical-effect threshold. All $n$ observations remain in the denominator ($n=30$ in the screen, $n=100$ in the confirmation), with ties counted as failures. Under this model, the indicator counts $K_r,K_d$ give exact one-sided binomial component tests \citep{LehmannRomano2005}:
\begin{equation}
 p_x=\sum_{k=K_x}^{n}{n\choose k}2^{-n},\qquad p_c=\max(p_r,p_d).\label{eq:sign}
\end{equation}
This is an intersection--union test (IUT) \citep{Berger1982IUT}. Under the union null that at least one probability is at most $1/2$, $p_c$ is valid whenever the component tests are valid: if component $x$ is null, $\Pr(p_c\le t)\le\Pr(p_x\le t)\le t$. This requires neither independence of the two endpoints nor symmetry of their difference distributions. It tests two marginal directional probabilities, not the probability of their joint occurrence.

The screening analysis applies Holm at familywise level $0.05$ to all 540 replication conditions \citep{Holm1979}, which allows arbitrary dependence between conditions; no effect-size filter defines the rejection set. Benjamini--Yekutieli (BY) and Benjamini--Hochberg (BH) adjustments are reported as false-discovery-rate (FDR) sensitivities, the latter conditional on dependence assumptions not established here \citep{BenjaminiYekutieli2001}. Discovery is analyzed separately. This procedure and the numerical tolerance were chosen after the screening results were known, so their error guarantees do not cover that choice. Effect sizes are reported separately against the pre-specified thresholds of 10\% median MSE improvement and $\tau_R=0.039962$ normalized regret.

\subsection{Screening sampler}\label{sec:streams}
The screening sampler hashes each point/scale/repetition label into a 31-bit seed and initializes a new generator at each point. In the four conditions retained by Holm, this produces 637--1059 colliding seed pairs between repetitions and 16--51 within repetitions per condition, all at distinct grid points (Appendix~\ref{app:statistics}). The screening repetitions therefore do not implement the independent-sampling model exactly, and Holm's robustness to dependence between conditions does not extend to dependence within a component test. The four screening rejections are accordingly model-based and exploratory.

\subsection{Prospective targeted confirmation}\label{sec:confirmation_design}
A protocol fixed and hashed before any new sample was drawn specifies the four screening candidates, $n=100$, the screening grids, circuits, exact ideal references and integer allocations; the confirmation uses only new samples. A new 256-bit entropy value initializes NumPy \texttt{SeedSequence}; separately spawned \texttt{PCG64DXSM} streams are consumed sequentially across grid points for each group, repetition and method/scale.

The primary indicators remain $r_{\MSE}>0$ and $d_R>10^{-12}$, and Eq.~\eqref{eq:sign} uses all 100 repetitions. Holm at $.05$ is applied to the four intersection--union p-values, with no exclusions or interim analyses; conclusions are conditional on the selection from the screen. Appendix~\ref{app:confirmation} gives allocations and design sensitivity.

\subsection{Computational verification}\label{sec:verification}
A second, separately written implementation reconstructs all 400 paired confirmation endpoints from the saved landscapes, evaluates binomial tails with integer binomial coefficients and recomputes the Holm adjustments.

\section{Results}
\subsection{Fresh confirmation of four selected conditions}
All four selected conditions pass Holm at $.05$ on 100 fresh repetitions (Table~\ref{tab:confirmation}). Median relative MSE improvements range from 7.65\% to 73.09\%, and median normalized regret shifts from 0.00308 to 0.00732, or 0.0109--0.0260 expected-cut units. Three of the four MSE medians exceed the pre-specified 10\% threshold, the exception being H3 quadratic at 7.65\%, and no regret median reaches the pre-specified threshold $\tau_R$: the directional effects are clear, while their practical size is small.

\begin{table}[htbp]
\caption{Prospective targeted family, $n=100$ per condition. $K_r$ and $K_d$ count positive indicators; $r$ and $d_R$ are medians. Native shift is $D_0\operatorname{median}(d_R)$ in expected-cut units. All four conditions were fixed before new sampling; Holm corrects these four tests only.}
\label{tab:confirmation}\centering\small
\setlength{\tabcolsep}{2.5pt}
\begin{tabular}{llrrrrrr}\hline
Graph & $(p,B,m)$ & $K_r$ & $K_d$ & $r$ & $d_R$ & \shortstack{Native\\shift} & Holm $p$\\\hline
H1 & $(0.02,1024,2)$ & 100 & 66 & 0.4358 & 0.00646 & 0.01520 & $8.95\times10^{-4}$\\
H3 & $(0.02,1024,2)$ & 100 & 70 & 0.0765 & 0.00643 & 0.01361 & $1.18\times10^{-4}$\\
H8 & $(0.01,4096,3)$ & 100 & 84 & 0.2613 & 0.00732 & 0.02596 & $5.21\times10^{-12}$\\
H8 & $(0.01,4096,2)$ & 100 & 67 & 0.7309 & 0.00308 & 0.01093 & $8.74\times10^{-4}$\\
\hline\end{tabular}
\end{table}

\subsection{Exploratory screen}
\begin{figure}[htbp]\centering
\includegraphics[width=0.8\textwidth]{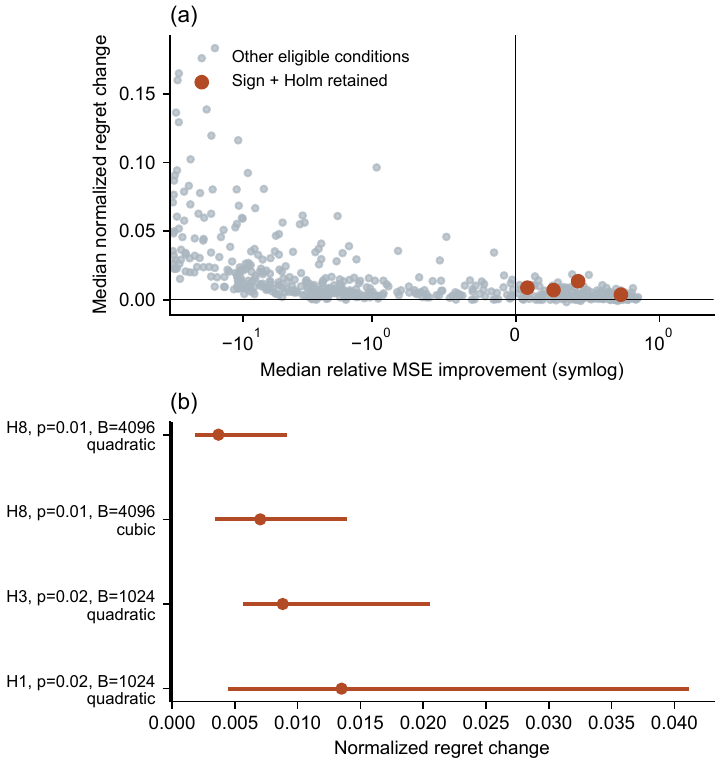}
\caption{Paired effects in the replication family. (a) Median relative mean-squared-error improvement and median normalized regret change for the 507 eligible conditions; excluded conditions are omitted, and the four Holm rejections are shown as larger orange points. The horizontal axis uses a symmetric logarithmic scale to retain large negative relative improvements. (b) Medians and seed interquartile ranges of the regret change for the four retained conditions.}
\label{fig:revised}
\end{figure}

Four of 540 replication conditions meet the Holm cutoff, distributed across H1, H3 and H8 (Table~\ref{tab:revised}). All four improve MSE in 30 of 30 repetitions; increased regret beyond numerical tolerance occurs in 26 or 28 repetitions. Their median MSE improvements range from 8.05\% to 73.11\%; three meet the pre-specified 10\% threshold, and the H3 quadratic condition does not. The median native shifts range from 0.0131 to 0.0317 expected-cut units, or 0.146\%--0.635\% of total graph edge weight, and none reaches $\tau_R$. Figure~\ref{fig:revised} places them among all eligible conditions and displays within-condition variability; its interquartile bars are descriptive.

\begin{table}[htbp]
\caption{Four replication conditions retained by the sign/IUT/Holm analysis of the screen. Columns as in Table~\ref{tab:confirmation}, with counts out of 30 repetitions.}
\label{tab:revised}\centering\small
\setlength{\tabcolsep}{3.5pt}
\begin{tabular}{llrrrrr}
\hline
Graph & $(p,B,m)$ & $K_d$ & $r$ & $d_R$ & Native shift & Holm $p$\\\hline
H1 & $(0.02,1024,2)$ & 26 & 0.4338 & 0.01350 & 0.03174 & 0.015999\\
H3 & $(0.02,1024,2)$ & 28 & 0.0805 & 0.00881 & 0.01865 & 0.000234\\
H8 & $(0.01,4096,3)$ & 26 & 0.2627 & 0.00703 & 0.02491 & 0.015999\\
H8 & $(0.01,4096,2)$ & 28 & 0.7311 & 0.00370 & 0.01311 & 0.000234\\
\hline
\end{tabular}
\end{table}

Holm and BY retain the same four replication conditions, while BH retains twelve; these screening results are exploratory (Section~\ref{sec:streams}). The retained sets are unchanged at regret tolerances $10^{-14}$, $10^{-12}$, $10^{-10}$ and $10^{-6}$. In the separately analyzed discovery family, Holm retains one condition, in G2. Appendix~\ref{app:statistics} reports the complete sensitivities, including strict floating-point positivity and the original bootstrap classification.

\subsection{Bounded bias--variance control}\label{sec:control}
An explanatory control uses all three discovery graphs, two endpoint noise parameters ($0.001,0.02$), $B=1024$, all three orders and the unmitigated baseline. It evaluates a new $12\times12$ grid and 500 independent multinomial landscape repetitions per configuration, and complements rather than retests the 540-condition endpoint (Appendix~\ref{app:control}).

Exact density-matrix probabilities give each scale's mean $\mu_j(\theta)$ and single-shot variance $v_j(\theta)$. With $b_Z=\sum_j\lambda_j\mu_j-L_0$, the exact decomposition is
\begin{equation}
 \E\!\left[\MSE(\widehat L_Z,L_0)\right]
 =\frac1{|\Theta|}\sum_{\theta}\left[b_Z(\theta)^2+
 \sum_j\lambda_j^2v_j(\theta)/B_j\right].\label{eq:biasvariance}
\end{equation}
At $p=.02$, linear and quadratic ZNE reduce expected MSE on all three graphs; cubic ZNE increases it despite reducing squared bias. At $p=.001$, all three orders increase expected MSE. The Monte Carlo mean MSE agrees with the exact decomposition within 1.919 Monte Carlo standard errors in every configuration.

\begin{figure}[htbp]\centering
\includegraphics[width=0.85\textwidth]{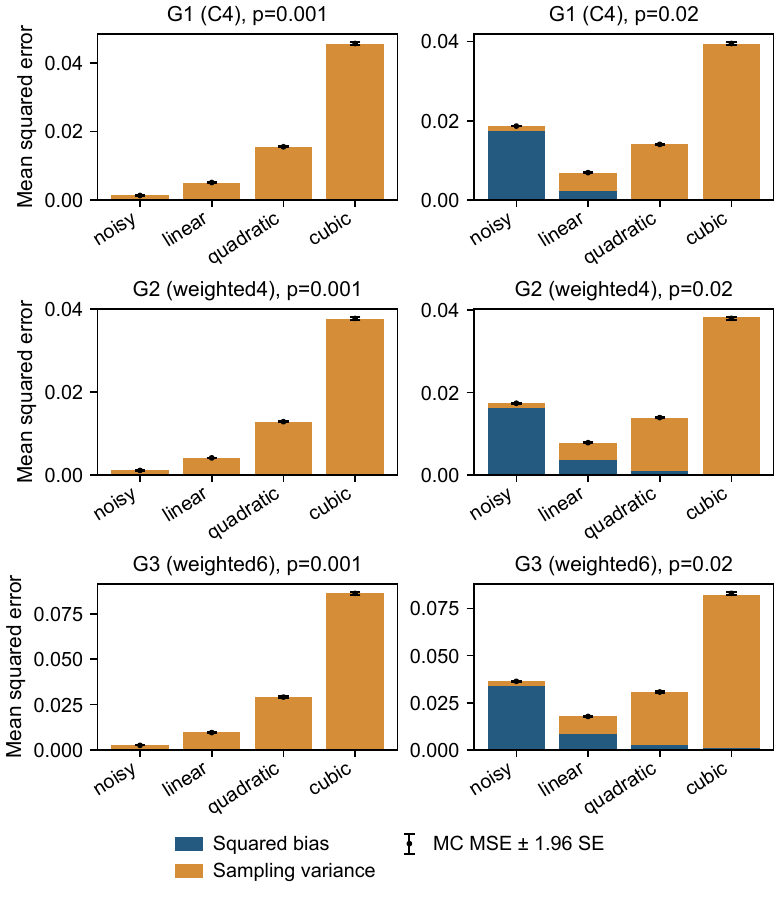}
\caption{Bias--variance control on a $12\times12$ grid with $B=1024$ shots per point for every method and 500 repetitions per configuration: exact grid-averaged squared bias and sampling variance (stacked bars) and Monte Carlo estimates of the mean-squared error with $\pm1.96$ standard errors (points). This control is separate from the $120\times120$ screen.}
\label{fig:biasvariance}
\end{figure}

\begin{figure}[htbp]\centering
\includegraphics[width=0.85\textwidth]{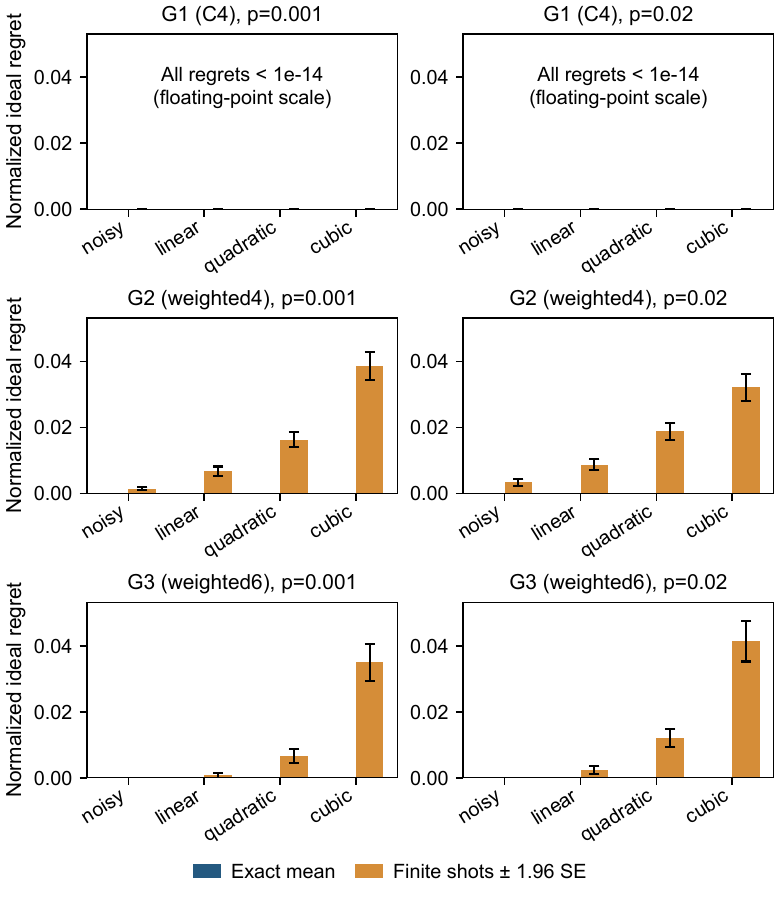}
\caption{Decision outcomes for the bias--variance control ($12\times12$ grid, $B=1024$, 500 repetitions per configuration): normalized ideal regret of exact-mean and finite-shot selections. A common vertical scale is used for all graphs. Error bars denote $\pm1.96$ Monte Carlo standard errors and are descriptive.}
\label{fig:control_decision}
\end{figure}

All exact mean landscapes select ideal minima to numerical precision. Nevertheless, finite-shot ZNE selections have greater Monte Carlo mean regret on G2 and G3 in the tested regimes. For G2 at $p=.02$, expected MSE falls from $0.017405$ without mitigation to $0.007840$ with linear ZNE, while Monte Carlo mean normalized regret increases from $0.003365$ to $0.008734$. Quadratic ZNE gives MSE $0.013914$ and Monte Carlo mean regret $0.01883$. In each G1 configuration, all 500 repetitions select symmetry-related ideal minima to numerical precision on this coarse grid (Figure~\ref{fig:control_decision}).

This control separates contributions to expected MSE exactly. Since the mean selections are ideal, the bias route of Section~\ref{sec:routes} alone produces no regret here, and the regret observed after sampling arises through the variance route. Residual bias can still change the margins that sampling must overcome, so the control identifies the route but not the separate causal contributions of bias and variance.

\subsection{Sensitivity to shot allocation}
An additional exact-moment analysis compares the implemented allocation with a pointwise oracle that knows every scale variance and minimizes ZNE variance over positive integer shot counts at the same total per-point budget. Across the four confirmation conditions, this reduces grid-mean variance by 0.0443\%--0.3652\%; even the largest pointwise improvement allowed by the continuous optimum is below 1.26\%. Appendix~\ref{app:allocation} gives each condition and a fixed-allocation comparison. The oracle uses the saved probabilities and is charged no calibration cost; the comparison concerns variance only.

\section{Discussion and limitations}\label{sec:discussion}
In the four confirmed conditions, ZNE improves landscape MSE in every fresh repetition, while the directional tests indicate a probability above one half of increased regret. These decision losses are small but reproducible and are not captured by aggregate MSE. In the control, ideal mean selections become imperfect under finite-shot sampling. The bound in Appendix~\ref{app:mathematics} relates this variance route to the mean margins in Eq.~\eqref{eq:contrast} and the noise scale $W_G\sqrt{S_B}$, where $W_G$ is the total edge weight. Near-optimal competitors with small margins can remain sensitive to sampling even when aggregate MSE falls. Richardson weights increase the variance-bound coefficient relative to unmitigated estimation at the same budget, while residual bias also changes those margins. This provides a structural interpretation of the small observed regret shifts.

Relative to prior work, the contribution is a paired measurement. Bias--variance trade-offs in Richardson extrapolation, mitigation-induced landscape roughness and the accuracy--decision distinction are known from theory, landscape diagnostics and optimizer runs \citep{Krebsbach2022RichardsonOptimization,Wang2024Trainability,Liu2023OSCAR,SackEgger2024MLMitigation}. Here the same sampled landscape supplies both whole-landscape MSE and ideal regret at an equal shot budget, which isolates the selection step from optimizer dynamics; a prospective confirmation shows that the resulting discordance reproduces on fresh streams; and exact moments show the variance route at work in a bounded control. The related decision-kernel preprint develops residual-gap theory and evaluates failure probabilities and dynamic regret \citep{Scavino2026DecisionKernels}; the present work supplies the corresponding empirical characterization for conventional Richardson extrapolation on exhaustive grids. The finite-grid bounds used here are elementary and serve to organize this evidence.

Several restrictions matter. The primary circuits are shallow and small, with local depolarizing noise. The generated graphs satisfy stratification and grid-adequacy rules, so the observed frequency is not a prevalence estimate for arbitrary QAOA instances. The ideal oracle and exhaustive grid are simulation tools; an adaptive optimizer on hardware has different resource and selection constraints. Measurement-frugal variational optimizers can instead allocate shots by iteration and gradient component \citep{Kubler2020MeasurementFrugal}; the present benchmark uses a prescribed per-point allocation. Auxiliary exploratory depth-two slices, thermal/readout simulations and COBYLA checks are summarized in Appendix~\ref{app:auxiliary}; the COBYLA conditions did not meet their prescribed endpoint-effect rule.

\section{Conclusion}
In this benchmark, accuracy gains from ZNE seldom came with decision losses that survive strict testing: a model-based familywise analysis retains four of 540 replication conditions. Those four losses are nevertheless reproducible. A frozen confirmation on fresh random streams recovered all of them, with median MSE improvements of up to 73.09\% alongside median normalized regret increases of 0.0031--0.0073. The losses are small, but aggregate MSE does not reveal them. Exact moments on a bounded control show the variance route producing such losses: mitigated mean landscapes keep the ideal minima, and the regret appears only once finite-shot landscapes are sampled and minimized, a step that a grid-averaged error does not resolve.

When mitigation is assessed as a component of optimization, decision regret, or a validated decision-level proxy, should therefore be reported alongside estimation error; the two can disagree even when the shot allocation is close to variance-optimal. Extending paired accuracy--decision measurements to deeper circuits, optimizer loops and hardware noise is the natural next step. Validated decision-level proxies will be needed in settings where the ideal reference is unavailable.

\section*{Acknowledgements}
The author received no specific funding for this work and declares no competing interests. OpenAI Codex and GPT-6.1-sol were used to assist with code development and review, statistical computation, checking of mathematical derivations, implementation of the control and confirmation analyses, and manuscript editing. AI tools are not authors. The author is responsible for the analyses and for the content of the manuscript.

\section*{Data availability}
Code and data supporting this study are available in the \href{https://github.com/vicenzoscavino1999/qaoa-zne-decision-regret}{GitHub repository} and archived on Zenodo (\href{https://doi.org/10.5281/zenodo.23607395}{doi:10.5281/zenodo.23607395}; version 1.0.0). The archive contains seed-level metrics, reference landscapes, complete targeted-confirmation data and protocol, analysis code and figure-generation scripts, with instructions for reproducing the reported results.

\begin{appendices}
\numberwithin{equation}{section}
\numberwithin{table}{section}
\section{Finite-grid mathematical details}\label{app:mathematics}
This appendix collects the finite-grid results referred to in Section~\ref{sec:mse_decisions}. The integer-budget bound formalizes the variance route of Section~\ref{sec:routes}, and the exact Richardson example illustrates the bias route.

\noindent\textbf{Integer-budget bound with an ideal optimal set.}
Let $\Theta_*=\{\theta:L_0(\theta)=L_*\}$ and choose a deterministic
$u\in\Theta_*$ independently of sampled landscapes. For the implemented integer
allocations $B_j\geq1$, write $b_p^{[a_j]}(\theta)=\E[\widehat L^{[a_j]}(\theta)]-L_0(\theta)$, $\eta_Z=\widehat L_Z-\E[\widehat L_Z]$, and define
\[
 S_B=\sum_j\frac{\lambda_j^2}{B_j},\quad
 b_Z(\theta)=\sum_j\lambda_j b_p^{[a_j]}(\theta),\quad
 M_u(\theta)=L_0(\theta)-L_*+b_Z(\theta)-b_Z(u).
\]
Assume independent shot samples across scales and grid points and a single-shot
objective interval width at most $W_G$. Hoeffding's inequality \citep{Hoeffding1963} gives, for every $\epsilon\geq0$,
\[
 \Pr\{R(\widehat L_Z;L_0)>\epsilon\}\leq
 \min\left\{1,\sum_{\theta:L_0(\theta)-L_*>\epsilon D_0}
 \exp\left[-\frac{(M_u(\theta)_+)^2}{W_G^2 S_B}\right]\right\}.
\]
\textit{Proof.} Here $x_+=\max(x,0)$. Each weighted individual shot has interval width $|\lambda_j|W_G/B_j$. Summing squared widths over the two independently sampled points gives $2W_G^2S_B$, so Hoeffding's inequality bounds the probability that $\eta_Z(\theta)-\eta_Z(u)\leq-M_u(\theta)$ by the displayed exponential when $M_u(\theta)>0$; otherwise the bound one is trivial.
If excessive regret occurs, the selected point is a summation point and its observed
loss is no larger than the observed loss at $u$. A union bound proves the claim.
At $\epsilon=0$ the bound concerns selection outside the entire ideal optimal set.
The comparison events may be dependent through their common anchor.
For the positive-weight graphs used here, single-shot objective values lie in
$[-W_G,0]$ with $W_G=\sum_e w_e$.
The same coefficient $S_B$ controls the variance bound in Eq.~\eqref{eq:integer_variance}.
The implemented largest-remainder allocations resolve ties between equal fractional
parts in floating point, which can differ from exact-rational resolution;
reproduction must therefore retain the recorded $B_j$.

This bound uses exact mean residuals and fixed independent sampling budgets.
Replacing the residuals by estimates obtained from the same repetitions does not
produce a certified probability bound. The bound concerns the optimal-set event
rather than a change of argmin coordinate, because a coordinate change between
equivalent minima has zero ideal regret. No empirical tightness or
predictive-accuracy claim is made for it.

\noindent\textbf{Exact abstract Richardson example.}
Consider deterministic expectation landscapes on $\Theta=\{0,1\}$ with
$L_0(0)=0$, $L_0(1)=\Delta>0$. For scales $a\in\{1,3\}$ define
\[
 b_p^{[a]}(0)=acp,\qquad b_p^{[a]}(1)=acp+a^2hp^2,
 \qquad c,h>0.
\]
With Richardson weights $(3/2,-1/2)$, the ZNE bias is $(0,-3hp^2)$.
The raw expectation landscape always preserves the ideal decision, while ZNE
reverses it strictly when $p>\sqrt{\Delta/(3h)}$. The exact deterministic MSEs are
\[
 \MSE(L_p^{[1]},L_0)=c^2p^2+chp^3+\frac{h^2p^4}{2},\qquad
 \MSE(L_Z,L_0)=\frac{9h^2p^4}{2}.
\]
Consequently ZNE has lower MSE exactly for
\[
 0<p<\frac{c(1+\sqrt{17})}{8h}.
\]
For any reference $p_0>0$, set $\Delta=1$, $h=p_0^{-2}$ and $c=3p_0^{-1}$.
Both strict reversal and strict MSE improvement hold throughout
$p_0/\sqrt3<p\leq p_0$. At $p=p_0$, the raw and ZNE MSEs are $12.5$ and $4.5$,
and their regrets are zero and one. A common shift by $-20$ places all ideal and
folded expectation values in $[-20,0]$ on $0\leq p\leq p_0$.
This is an exact abstract bounded-observable example, not a model of the studied
circuits; its deterministic MSEs exclude shot variance, which the integer-budget
bound treats.

\section{Statistical methods and conventions}\label{app:statistics}
C denotes the discovery family and D the generated replication family in the tables below.
\subsection{Pre-specified screening rule}
The screen was planned with a bootstrap classification rule based on 10\,000 percentile-bootstrap resamples of the 30 paired seed metrics \citep{EfronTibshirani1993}. A nominal condition required median $r_{\MSE}\ge0.10$, a positive lower bootstrap limit for $r_{\MSE}$, positive median $d_R$, nonnegative lower bootstrap limit for $d_R$, and at least 60\% positive $d_R$ values. Bootstrap positivity tails with a one-count correction were combined using their maximum, followed by BH adjustment \citep{BenjaminiHochberg1995}, with $q=0.10$ for discovery and $q=0.05$ for replication. Nine discovery and 33 replication conditions were excluded because at least one seed had noisy MSE below $10^{-4}D_0^2$. No condition triggered the extrapolation-breakdown rule, which required more than half the grid to fall outside the ideal range extended by $0.10D_0$.

Bootstrap positivity tails need not be null-calibrated p-values, so this rule does not yield calibrated familywise statements. Its zero Holm rejections also reflect a Monte Carlo resolution limit, examined in Appendix~\ref{app:bootstrap_resolution}. Under the rule, 18 discovery conditions are nominal and 13 survive BH, and 69 replication conditions are nominal and 39 survive BH. We report these classifications as a secondary descriptive analysis.

\subsection{Sampling unit and complete families}
Every screening condition contains the same 30 seed labels. The reference test model assumes independent, identically distributed shot repetitions conditional on a fixed graph, noise model, circuit folding, integer allocation and grid; Appendix~\ref{app:collisions} describes how the screening sampler departs from it. Metrics are formed from the paired noisy and mitigated landscapes. Sharing the noisy baseline between orders creates cross-condition dependence, but does not require assuming independent endpoints within a pair for an intersection--union test.

Discovery and replication are separate families of 135 and 540 tests. Conditions excluded by the detection-floor rule receive adjusted input p-values of one, so they remain nonrejectable members of their families. An exclusion based on observed data can only increase these p-values; it cannot invalidate super-uniformity of an otherwise valid unfiltered test.

\subsection{Exact directional tests and multiplicity}
For $X=r_{\MSE}$ and $Y=d_R$, the component nulls are $\Pr(X>0)\le1/2$ and $\Pr(Y>10^{-12})\le1/2$. Under that sampling model, standard exact binomial testing \citep{LehmannRomano2005} gives, for either positive count $K$, the p-value $\sum_{k=K}^{30}\binom{30}{k}2^{-30}$. A binomial upper tail is largest at the boundary $1/2$ of this composite null. Zeros and numerical ties are failures, with denominator 30; deleting them would change the estimand.

The conjunction alternative requires both marginal probabilities to exceed one half; this is the intersection--union construction \citep{Berger1982IUT}, whose validity is shown in Section~\ref{sec:directional}. Holm's sorted adjusted values \citep{Holm1979} are
\[
 p^{\rm Holm}_{(i)}=\min\{1,\max_{l\le i}(T-l+1)p_{(l)}\},
\]
where $T$ is the number of tests in the family; they control the familywise error at level $\alpha$ under arbitrary dependence between conditions. This is distinct from an FDR claim on an outcome-filtered subset. BY sensitivity \citep{BenjaminiYekutieli2001} uses the harmonic factor $\sum_{l=1}^T1/l$. BH sensitivity \citep{BenjaminiHochberg1995} does not supply arbitrary-dependence control.

\subsection{Seed collisions in the screening sampler}\label{app:collisions}
The screening code initializes \texttt{default\_rng} at every point and scale from the first eight hexadecimal SHA-256 digits masked with \texttt{0x7fffffff} (31 bits). Distinct labels can therefore reuse a stream. Counting all random-number-generator (RNG) initializations in each of the four conditions retained by Holm gives:
\begin{center}\small
\begin{tabular}{lrrr}\hline
Condition $(G,p,B,m)$ & RNG calls & Within repetition & Between repetitions\\\hline
H1, .02, 1024, 2 & 1\,728\,000 & 16 & 677\\
H3, .02, 1024, 2 & 1\,728\,000 & 32 & 690\\
H8, .01, 4096, 3 & 2\,160\,000 & 51 & 1059\\
H8, .01, 4096, 2 & 1\,728\,000 & 28 & 637\\\hline
\end{tabular}
\end{center}
The last two columns count colliding pairs of RNG initializations, not duplicated landscapes or changed decisions, and all colliding pairs refer to distinct grid indices. The counts therefore do not quantify the effect on signs; the directional analysis retains the four conditions. Repeated streams nevertheless induce dependence not represented by the ideal binomial model, so the four screening results are model-based diagnostics, without a certified error probability for this generator.

\subsection{Roundoff-sensitive signs}
Strict floating-point positivity, without the numerical tolerance, retains five D conditions. In H3 at $p=.02$, $B=4096$, cubic order, two positive $d_R$ values are approximately $2.1\times10^{-16}$. Discounting them as numerical ties lowers its regret count from 26 to 24, and it no longer survives Holm. H8 at $p=.01$, $B=4096$, cubic order also has two roundoff-level positives; its count falls from 28 to 26, and it remains retained.

\begin{table}[htbp]\centering\small
\caption{Directional-test sensitivity to the regret threshold. Holm uses $.05$ for each separate family. BY/BH use $.10$ for C and $.05$ for D. Values are counts before any practical-effect filtering.}
\begin{tabular}{lrrrrrr}\hline
Threshold & C Holm & C BY & C BH & D Holm & D BY & D BH\\\hline
$0$ & 2 & 2 & 11 & 5 & 5 & 34\\
$10^{-14}$ & 1 & 0 & 4 & 4 & 4 & 12\\
$10^{-12}$ & 1 & 0 & 4 & 4 & 4 & 12\\
$10^{-10}$ & 1 & 0 & 4 & 4 & 4 & 12\\
$10^{-6}$ & 1 & 0 & 4 & 4 & 4 & 12\\\hline
\end{tabular}
\end{table}
The tolerance is in normalized ideal-regret units, far below the observed median shifts. The exact median relative MSE improvement of the retained H3 quadratic condition is $0.080523$.

\begin{table}[htbp]
\caption{Pre-specified bootstrap classifications and directional tests. Bootstrap rows include the pre-specified nominal filters; sign rows do not. Consequently, these rows differ in estimand and decision rule, not only in multiplicity adjustment. C and D are separate families.}
\label{tab:sensitivity}\centering\small
\begin{tabular}{lrrr}\hline
Analysis & Holm $.05$ & BY & BH\\\hline
D: pre-specified bootstrap rule & 0 & 15 & 39\\
D: exact conditional bootstrap tail, same filters & 14 & 15 & 40\\
D: sign tests, strict floating-point positivity & 5 & 5 & 34\\
D: sign tests, $d_R>10^{-12}$ & 4 & 4 & 12\\
C: sign tests, $d_R>10^{-12}$ & 1 & 0 & 4\\\hline
\end{tabular}
\par\smallskip\footnotesize BY/BH use $.05$ for D and $.10$ for C. Exact conditional bootstrap tails remove Monte Carlo error only; they are not exact frequentist p-values.
\end{table}

\subsection{Isolating bootstrap resolution from calibration}\label{app:bootstrap_resolution}
For the empirical bootstrap \citep{EfronTibshirani1993}, the lower bound $1/(10000+1)$ on Monte Carlo bootstrap tails exceeds the first Holm cutoff $.05/540$. To isolate this resolution limit, we also compute the \emph{conditional empirical-bootstrap} median distribution without Monte Carlo sampling.

For $n=2s$ and empirical distinct values $x_i$ with probabilities $p_i$, write $F_i=\sum_{l\le i}p_l$ and $F_i^-=F_i-p_i$. If the two middle order statistics equal distinct values $x_i<x_l$, their joint probability is
\[
 {2s\choose s}\{F_i^s-(F_i^-)^s\}
 \{(1-F_l^-)^s-(1-F_l)^s\}.
\]
If both middle values equal $x_i$, the probability is
\[
 \sum_{a=0}^{s-1}\sum_{b=0}^{s-1}
 \frac{(2s)!}{a!b!(2s-a-b)!}
 (F_i^-)^a(1-F_i)^b p_i^{2s-a-b}.
\]
Assign these masses to $(x_i+x_l)/2$ and $x_i$, respectively. The implementation verifies normalization and compares the cumulative distribution function to exhaustive resampling enumeration for small samples with and without ties. The pre-specified nominal filters remain attached only to this resolution sensitivity: D gives 40 BH, 15 BY and 14 Holm classifications, compared with 39, 15 and zero under the Monte Carlo rule.

Sign-test rejection probabilities are also enumerated at the null boundary.

\subsection{Graph-level interpretation}
The pre-specified stress rule counts conditions with $p\in\{.005,.01,.02\}$, $B\in\{1024,4096\}$ and all orders, requiring at least three nominal classifications per graph. The observed count is 11/12. The six fixed strata have two selected graphs each, and the positivity probabilities need not be identical across strata. The homogeneous binomial tail $13/4096$ is not an exact test of a heterogeneous population-average probability. A related bound would require the stronger null that \emph{every} independent graph-specific probability is at most $1/2$. We report the 11/12 count descriptively.

\section{Prospective targeted confirmation}\label{app:confirmation}
\subsection{Frozen design and sampling}
The protocol was fixed before any confirmatory landscape was sampled. Its machine-readable form, explanatory text and SHA-256 lock identify the analysis script, graph definitions, exact ideal arrays and source modules; its timestamped record is included in the reproducibility package, and it is not externally registered. All four screening candidates were retained: H1 and H3 at $p=.02$, $B=1024$, quadratic order; H8 at $p=.01$, $B=4096$, quadratic and cubic orders. The planned sample is 100 repetitions for every condition, without pooling screening data.

In increasing scale order, the exact integer allocations are
\[
(548,366,110),\qquad (2194,1463,439),\qquad (1493,1493,896,214)
\]
for quadratic order at $B=1024$, quadratic order at $B=4096$ and cubic order at $B=4096$, respectively. The same $120\times120$ grids, row-major first-index argmin, circuit and depolarizing model are used. Probabilities must be finite, nonnegative and sum to one within $10^{-9}$; normalization by the sum removes floating-point drift before sampling. The ideal arrays match independent statevector calculations at eight fixed points per graph.

A recorded 256-bit entropy value seeds the root \texttt{SeedSequence}. Children are spawned in group order H1, H3, H8, then repetition 0--99, then baseline followed by the cell/scale order in the protocol. H8 cubic precedes quadratic. Each child initializes one \texttt{PCG64DXSM} generator, consumed sequentially over grid points. There are 1600 distinct stream paths and initial states; baseline draws are shared only between the two H8 orders at the same repetition. Distinct paths and states support the usual independent-stream model. Software versions, including NumPy 2.2.6, are fixed in the protocol.

A preliminary pilot that measured only runtime and pointwise agreement with Aer is excluded, together with its streams and draws. No interim endpoint analysis, optional stopping, sample-size extension or metric-based exclusion is permitted.

\subsection{Inference and complete outcomes}
Before execution, exact design calculations showed that 62 successes in both components suffice to pass the most stringent Holm cutoff $.0125$. For a component success probability $.60$, $.65$ or $.70$, the probability of reaching 62 is $.3822$, $.7699$ or $.9660$, respectively. These are design scenarios, not estimates from the screening success rates. Power can be limited for probabilities close to one half.

\begin{table}[htbp]\centering\small
\caption{Complete primary p-values for the targeted family ($n=100$ each). The final column is the descriptive count of repetitions with both positive indicators; it is not the tested probability.}
\setlength{\tabcolsep}{3pt}
\begin{tabular}{lrrrrr}\hline
Condition & $p_r$ & $p_d$ & $p_{\rm IUT}$ & $p_{\rm Holm}$ & Both\\\hline
H1, $m=2$ & $7.89\times10^{-31}$ & $8.95\times10^{-4}$ & $8.95\times10^{-4}$ & $8.95\times10^{-4}$ & 66\\
H3, $m=2$ & $7.89\times10^{-31}$ & $3.93\times10^{-5}$ & $3.93\times10^{-5}$ & $1.18\times10^{-4}$ & 70\\
H8, $m=3$ & $7.89\times10^{-31}$ & $1.30\times10^{-12}$ & $1.30\times10^{-12}$ & $5.21\times10^{-12}$ & 84\\
H8, $m=2$ & $7.89\times10^{-31}$ & $4.37\times10^{-4}$ & $4.37\times10^{-4}$ & $8.74\times10^{-4}$ & 67\\
\hline\end{tabular}\end{table}

Medians, quartiles, native regret shifts and individual 95\% Clopper--Pearson intervals accompany all 400 paired endpoints in the reproducibility package. The intervals are descriptive and do not have simultaneous family coverage. The verification implementation of Section~\ref{sec:verification} also checks the stream records, complemented by partial stream replay and bounded Aer checks.

Probability generation and sampling took 535.74 and 160.55 seconds, respectively (11.60 minutes combined), with probabilities cached and reused across repetitions; these timings indicate scale only.

\section{Bias--variance control: design and validation}\label{app:control}
The control of Section~\ref{sec:control} uses the three discovery graphs, chosen for coverage, and was designed after the screening results were known. Its 24 graph/noise/method configurations comprise 3456 pointwise rows and 12\,000 complete Monte Carlo landscape repetitions, with independent random streams across methods, scales and grid points; common repetition indices are labels, not common random numbers.

The simulator's moments agree with Aer at two specified validation cases to maximum discrepancies $2.73\times10^{-14}$ and $1.33\times10^{-13}$. An independent ideal statevector calculation at all 432 graph/grid points agrees within $6.22\times10^{-15}$.

The three ideal grids contain 32, two and two equivalent minima within $10^{-10}$ objective units. The saved outcomes also record an illustrative single-competitor contrast, between the first ideal minimum and the point with the smallest strictly positive ideal gap.

The control took 69.29 seconds, compared with approximately 10.21 and 50.46 hours for the discovery and replication screens on different machines.

\section{Allocation sensitivity}\label{app:allocation}
\label{sec:allocation_sensitivity}
This deterministic sensitivity analysis uses the saved probability caches of the four confirmation conditions and does not modify the confirmation. At each grid point, basis-state objective values $L_x$ and cached probabilities $p_{j,x}$ give
\[
 \mu_j=\sum_x p_{j,x}L_x,\qquad
 v_j=\sum_xp_{j,x}(L_x-\mu_j)^2.
\]
All scale variances in these caches are positive. For fixed weights, independent scale samples and total budget $B$, the continuous variance minimum is
\[
 V_{\rm cont}^*(\theta)=
 \frac{\bigl(\sum_j|\lambda_j|\sqrt{v_j(\theta)}\bigr)^2}{B},
 \qquad B_j^*(\theta)\ \propto\ |\lambda_j|\sqrt{v_j(\theta)}.
\]
The actual variance uses the implemented integer $B_j$. For the exact positive-integer minimum, write $q_j=\lambda_j^2v_j$, initialize every count at one and assign each remaining shot to the largest decrease $q_j/[B_j(B_j+1)]$. These decreases are monotonically diminishing for each scale, so the greedy sequence gives the global optimum of this separable integer problem. A final one-shot exchange check is performed at every point.

The pointwise oracle may also change allocations between grid points, so it is a favorable reference for variance reduction, not an experimentally available protocol at the same acquisition cost. With an overbar denoting a grid mean, Table~\ref{tab:allocation_sensitivity} reports $100(1-\overline V_{\rm int}^*/\overline V_{\rm used})$. These are ratios of mean variances, not averages of pointwise ratios. A single fixed allocation for each entire condition is also optimized, using grid-mean $q_j$ in the integer calculation.

\begin{table}[htbp]\centering\small
\caption{Exact-moment allocation sensitivity for the four confirmation conditions. Pointwise and fixed-oracle gains are percentage reductions in grid-mean variance; all allocations retain the same total shots per point.}
\label{tab:allocation_sensitivity}
\setlength{\tabcolsep}{5pt}
\begin{tabular}{lrrrrr}\hline
Condition & $B$ & $\overline V_{\rm used}$ & $\overline V_{\rm int}^*$ & \shortstack{Pointwise\\gain (\%)} & \shortstack{Fixed\\gain (\%)}\\\hline
H1, $m=2$ & 1024 & 0.01565629 & 0.01563791 & 0.11738 & 0.01077\\
H3, $m=2$ & 1024 & 0.01219100 & 0.01218560 & 0.04433 & 0.00005\\
H8, $m=3$ & 4096 & 0.02333666 & 0.02325143 & 0.36521 & 0.17148\\
H8, $m=2$ & 4096 & 0.00821450 & 0.00819093 & 0.28689 & 0.13492\\
\hline\end{tabular}\end{table}

The largest pointwise reduction relative to the continuous lower bound is 1.2566\%, and the reduction in grid-mean excess variance above the unmitigated baseline is below 0.38\% in every condition. A continuous uniform allocation has greater grid-mean variance than the implemented allocation in all four conditions. These results concern these cached distributions and fixed estimator coefficients.

Because moments alone do not determine the selected minimum, the confirmation applies to its prescribed allocation.

\section{Auxiliary exploratory checks}\label{app:auxiliary}
The following exploratory checks accompany the screen. They are not pooled into the depth-one test family; their full outputs are archived with the screening data.

\begin{table}[htbp]\centering\small
\caption{Scope of the auxiliary checks. None supplies hardware or optimizer-loop confirmation of the paired endpoint.}
\begin{tabular}{L{.22\textwidth}L{.68\textwidth}}\hline
Diagnostic & Interpretation\\\hline
Depth-two slices & Selected instances, slices and budgets provide exploratory directional evidence only; they are not a full two-layer parameter search or an independent replication of the paired endpoint.\\
Thermal/readout noise & Synthetic component checks are not hardware-calibrated. The MSE-improvement precondition often fails, especially for thermal-only noise.\\
COBYLA endpoints & No examined instance/order condition met the prescribed joint rule of at least 80\% positive repetitions and median optimizer regret shift above .02.\\
No-reference scores & On the pre-specified D classifications, areas under the receiver-operating-characteristic curve were .546 for order-to-order argmin distance, .498 for seed instability and .504 for observed-margin risk. They do not validate an oracle-free replacement for ideal regret.\\
Leave-one-out checks & Leave-one-seed, graph and factor checks used the pre-specified bootstrap classifications. Their conclusions are not transferred to the sign/Holm set.\\\hline
\end{tabular}
\end{table}

For example, the H11 quadratic random-start COBYLA check had 22/25 positive repetitions and median shift about .0195, below its .02 criterion. The warm-start H11 quadratic check had a slightly negative median. These mixed results delimit the grid-based claim.

\end{appendices}
\bibliographystyle{plainnat}
\setlength{\bibsep}{0pt}
\interlinepenalty=10000
\bibliography{references}
\end{document}